\documentclass[preprint, onecolumn, secnumarabic, nobalancelastpage, amsmath, superscriptaddress, notitlepage]{revtex4-2}

\usepackage[T1]{fontenc}
\usepackage{color}
\usepackage{amsmath}
\usepackage[utf8]{inputenc}
\usepackage{amstext}
\usepackage{amssymb}
\usepackage{graphicx}
\usepackage{wasysym} 
\usepackage{nicefrac}
\usepackage{multirow}
\usepackage{caption}
\usepackage{subcaption}
\usepackage[justification=raggedright,singlelinecheck=false]{caption}
\newcommand{\RNum}[1]{\uppercase\expandafter{\romannumeral #1\relax}}
\makeatletter

\begin{document}
\title{Persistence length of short homopolymeric single-stranded DNA sequences in polyvalent cations}

\author{Balaka Mondal}
\affiliation{Department of Chemistry, The University of Texas at Austin, Austin, TX 78712, USA}

\author{D.~Thirumalai}
\affiliation{Department of Chemistry, The University of Texas at Austin, Austin, TX 78712, USA}
\affiliation{Department of Physics, The University of Texas at Austin, Austin, TX 78712, USA}

\email{dave.thirumalai@gmail.com}
\date{\today}

\begin{abstract}
We used simulations of short single stranded DNA (ssDNA) homopolymers, based on the sequence dependent Three Interaction Site (TIS) model, to calculate the persistence length ($l_p$) in polyvalent cations.  The TIS model accounts for stacking interactions and electrostatic interactions are treated using the Coulomb potential. We find that $l_p$ for $\mathrm{dT_{30}}$ (T is thymine) and $\mathrm{dA_{30}}$ (A is adenine) is quantitatively fit using $l_p = l_p^0 + \lambda \kappa^{-1}$ ($l_p^0$ is the bare persistence length, $\lambda$ is a dimensionless constant, and $\kappa$ is the inverse Debye length) in the divalent cations $\mathrm{Mg^{2+}}$ and $\mathrm{Ca^{2+}}$.  The dependence of $l_p$ on $\kappa$ is surprising because it was derived for long flexible polyelectrolytes in which the charges interact via the Debye-H\"{u}ckel potential.  The $l_p^0$ values are 0.4 nm and 1.1 nm for polyT and polyA, respectively. Strikingly, $l_p$ is almost independent of the tetravalent spermine concentration. There is  no clear theoretical explanation although simulations suggest that the number of spermine molecules that bind to the ssDNA saturates at a small value.  A qualitative picture, based on the restrictions of access to the phosphate groups due to volume exclusion of the anisotropic structure of $\mathrm{Spm^{4+}}$, rationalizes the simulation results.   The predicted dependence of $l_p$ in spermine awaits experimental test.  
\end{abstract}

\maketitle

\section*{Introduction}
Biological molecules, such as DNA and RNA, are highly charged.  In  general, including polyelectrolytes (referred to as PE), the persistence length ($l_p$)  is a measure of their sizes. However, the dependence of $l_p$ on the ionic concentration is difficult to calculate because of  the interplay of several length scales~\cite{Muthu23Book,Muthu17Macro}. If the PE is intrinsically stiff and if the interaction between
the charges is given by the   {Debye-H\"{u}ckel} potential, then the dependence of $l_p$ is well described by the 
Odijk-Skolnick-Fixman (OSF) theory \cite{Odijk1977,Skolnick1977}. The OSF theory shows that $l_p = l_{p}^{0} + l_{OSF}$, where $l_{p}^{0}$ is the bare
persistence length.  The electrostatic contribution, $l_{OSF}$, is given by,
\begin{equation}
l_{OSF} =  \frac{l_B}{4 \kappa^2 A^2}.
\label{OSF}
\end{equation}
In Eq. \ref{OSF}, the Bjerrum length, $l_B = \frac{e^2}{4 \pi \epsilon k_BT}$ ($e$ is the unit of charge, $\epsilon$ is the dielectric constant, $k_B$ is the Boltzmann constant, and $T$ is the temperature), $A$ is the
distance between the charges on the PE, and $\kappa$ is the inverse screening length ($\kappa^2 = 4 \pi l_B I$ where $I$ is the ionic strength). 

It is known that Eq. \ref{OSF} is valid when $A$, the distance between the charges on the PE is small ($A \ll (l_{p}^{0}l_B)^{1/2}$), which is  satisfied in double stranded DNA for instance. The situation is more complicated if the PE is intrinsically flexible ($A \gg (l_{p}^{0}l_B)^{1/2}$), as is the case in single stranded DNA (ssDNA) and RNA \cite{Tinland1997,Caliskan05PRL}.  Variational  theories \cite{Barrat93EL,Ha95Macro,Ha99JCP,Netz2003} have established that for flexible PEs, the electrostatic persistence length, $l_e$ scales as,
\begin{equation}
    l_p = l_{p}^{0} + l_e \sim (l_{p}^{0}l_B/A^2) \kappa^{-1}.
    \label{leFlex}
\end{equation}
Note that Eq. \ref{OSF} predicts that $l_e \sim I^{-1}$ whereas Eq. \ref{leFlex} shows that $l_e \sim I^{-.5}$. 

The assumption that Debye-H\"{u}ckel interactions between charged species, which is made in all the theories quoted above, is unlikely to be valid for divalent cations \cite{Thirumalai25COSB} and even less so for spermine, a tetravalent cation at neutral pH. 
In addition, for ssDNA, we expect that there ought to be differences between sequences, such as polyT and polyA because of differences in the stacking propensity.  To assess the applicability of the PE theories and to investigate sequence effects, we simulated ssDNA sequences $\mathrm{dT_{30}}$ and $\mathrm{dA_{30}}$ in $\mathrm{Mg^{2+}}$ and $\mathrm{Ca^{2+}}$ as well as in $\mathrm{Spm^{4+}}$ in the presence of 20 mM NaCl. We make two unexpected predictions. (1) The persistence length of the sequences varies as $l_p \sim l_p^0 + \lambda \kappa^{-1}$ ($l_p^0$ and the dimensionless constant $\lambda$ are dependent on sequence and $\kappa \sim \sqrt{I}$ where $I$ is the ionic strength). This result is surprising because the  variation,   $l_p \sim$   $\kappa^{-1}$, was derived for long flexible PE in which the electrostatic interactions were treated using the DH interaction, which are deemed to be inadequate for divalent cations \cite{Thirumalai25COSB,Denesyuk15NatChem}. 
(2) Strikingly, $l_p$ is almost independent   of $\mathrm{Spm^{4+}}$ concentration for both the sequences, which qualitatively differs from the behavior in divalent cations.  There is no theory for this observation although simulations suggest that, due to the size of  spermine, only a fixed number of $\mathrm{Spm^{4+}}$ cations bind to the phosphate groups at all concentrations. The prediction for $l_p$ on $\mathrm{Spm^{4+}}$ awaits future experiments.

\section*{Methods}

\begin{figure}[ht!]
\includegraphics[width=1.0\textwidth]{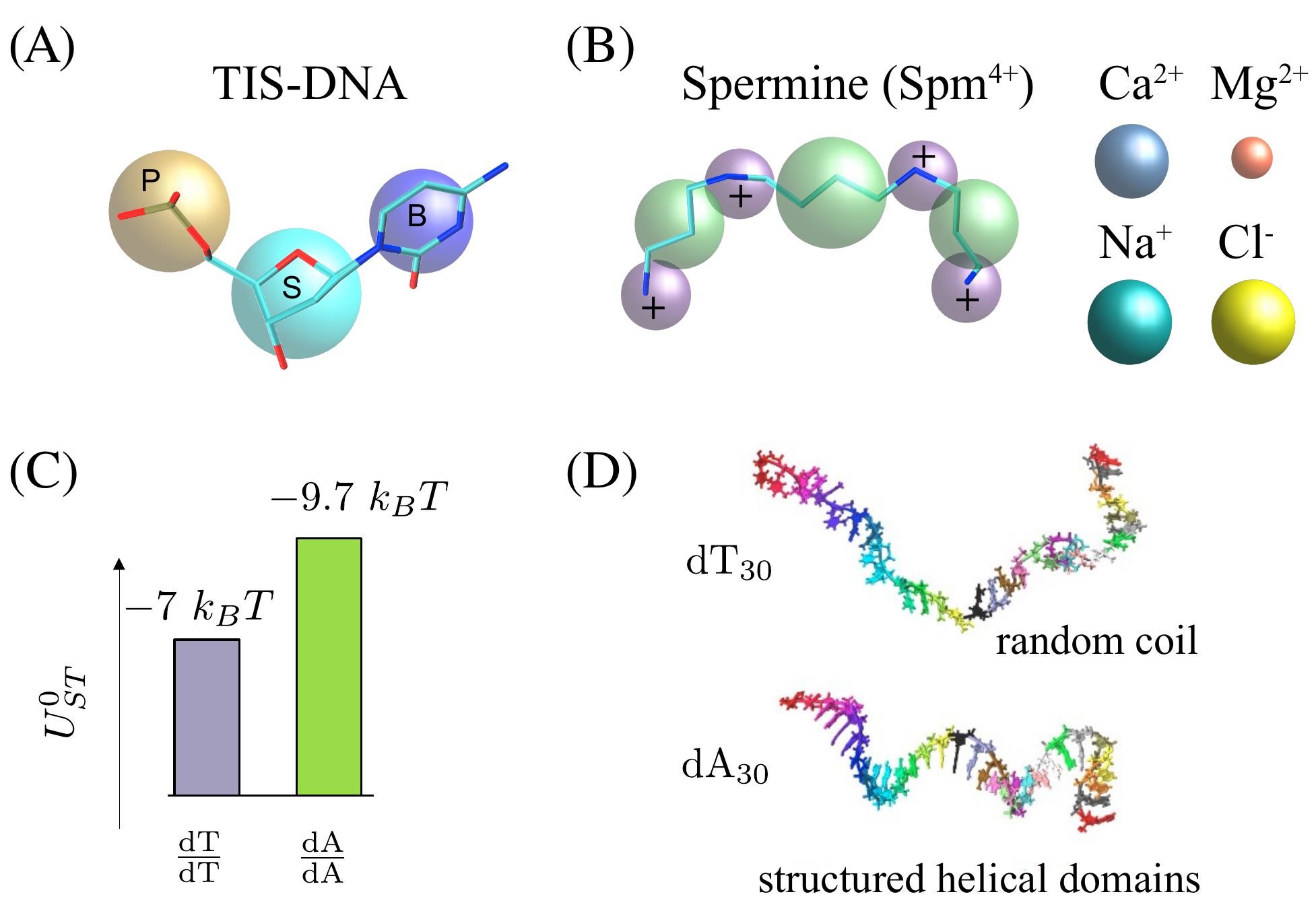}
	\caption{\label{fig:fig_0} 
\textbf{Three-interaction-site (TIS) model for DNA.}(A) Each nucleotide is represented by three beads, placed at the center of sugar (cyan), phosphate (tan) and base (blue) groups. (B) Monovalent ($\mathrm{Na^{+}}$, $\mathrm{Cl^{-}}$) and divalent ($\mathrm{Mg^{2+}}$, $\mathrm{Ca^{2+}}$) ions are modeled as single coarse-grained beads. The tetravalent $\mathrm{Spm^{4+}}$ is represented using seven coarse-grained beads. Purple beads represent charged amine groups and green beads correspond to alkyl groups. (C) Base stacking energies ($U_{ST}^{0}$) between thymine (dT) and adenine (dA) bases. (D) Random coil-like structure for $\mathrm{dT_{30}}$ and a helical conformation for $\mathrm{dA_{30}}$, reflecting the differences in the strength of the stacking interactions.
}
\end{figure}

\noindent\textbf{Three-Interaction-Site (TIS) model with explicit ions:} We used the three-interaction-site (TIS) coarse-grained model for nucleic acids~\cite{Hyeon05PNAS,Denesyuk15NatChem,Chakraborty18JCTC,Mondal24JCTC} in which  monovalent  $\mathrm{Na^{+}}$, divalent  ($\mathrm{Mg^{2+}}$ and $\mathrm{Ca^{2+}}$) and tetravalent cations $\mathrm{Spm^{4+}}$ are treated explicitly (see Figure~\ref{fig:fig_0}). In the TIS representation, each nucleotide is represented by three spherical beads placed at the center of mass of phosphate P,  sugar S and base B.  Ions are modeled as single spherical beads. The energy function for homopolymeric ssDNA is given by,
\begin{equation}
U_{TIS-ion} = U_{B} + U_{Ang} + U_{EV} + U_{St} + U_{El}. 
\end{equation}
The first four terms represent bond length and angle potentials, excluded volume interactions, and single-strand base stacking, respectively. The functional form for these interactions are given previously \cite{Chakraborty18JCTC,Mondal24JCTC}.

\noindent{\bf Electrostatic interactions, $U_{EL}$:} The interaction between two charged groups with charges $Q_i$ and $Q_j$, separated by a distance, $r_{ij}$, is given by the Coulomb potential,
\begin{equation}
U_{EL} = \frac{e^{2}}{4\pi\epsilon_{0}\epsilon(T)} \sum_{i,j} \frac{Q_{i}Q_{j}}{r_{ij}},
\end{equation}
\noindent where $\epsilon_{0}$ is the vacuum permittivity.  The temperature-dependent dielectric constant, $\epsilon(T) = 87.74 - 0.4008T + 9.398\times10^{-4}T^{2} - 1.410\times10^{-6}T^{3}$,  is taken from \cite{hasted1972water}, where the temperature, $T$, is in Celsius.
The charge on the phosphate group is $Q_P=-1$e (unit of electron charge).

\noindent\textbf{Model for $\mathrm{Spm^{4+}}$:}  We modeled $\mathrm{Spm^{4+}}$ using seven coarse-grained interaction centers (Figure~\ref{fig:fig_0}B).
Four sites represent the amine groups, and the remaining three are the alkyl segments.
The energy of a PA molecule is, $U_{PA} = U_{B} +  U_{EV} +  U_{El}$. 
The  bond stretch term, $U_{B}$, is taken to be a harmonic potential, $U_{B} = k_{b}(r-r_{0})^{2}$.  The equilibrium bond length, $r_{0}$,  is the sum of the radii of the bonded beads $i$ and $j$, and $k_{b}$ = 20.0 kcal/mol/\AA$^{2}$ is the spring constant. 
Excluded volume repulsion, $U_{EV}$, between any two non-bonded sites $i$ and $j$, separated by a distance, $r$ (in nm), is, 
\begin{equation}
U_{EV} = \sum_{i,j} \epsilon_{ij}\bigg[{\bigg(\frac{0.16}{r + 0.16 -D_{ij}} \bigg)}^{12} -   2{\bigg(\frac{0.16}{r + 0.16 -D_{ij}} \bigg)}^{6} + 1\bigg], \hspace{0.5in} r \leq D_{ij}, 
\label{eq:eq_ev}
\end{equation}
\noindent where $D_{ij} = R_{i} + R_{j}$ and $\epsilon_{ij}=\sqrt{\epsilon_{i}\epsilon_{j}}$;  $R_i$ and $R_j$ are the radii of $i^{th}$ and $j^{th}$ beads, respectively, and $\epsilon_{i}$ and $\epsilon_{j}$ are the respective interaction strengths.
The radius (=0.163 nm) and $\epsilon$ (=0.17 kcal/mol) values for the amine beads are taken from the  OPLS-AA \cite{Rizzo99JACS} force field parameters of amine nitrogen. We  used the same $\epsilon$ values for the amine and the alkyl groups because the short-range nature of the  repulsive excluded volume interaction should be insensitive to the $\epsilon$ values. The radii for the alkyl groups, $\mathrm{(CH_{2})_3}$ (=0.293 nm) and $\mathrm{(CH_{2})_4}$ (=0.308 nm), are taken from the Self-Organized Polymer model with side chains (SOP-SC) \cite{Liu11PNAS,Hyeon06STRUCTURE}, corresponding to the side chain radii of valine and leucine, respectively. The interactions between DNA and $\mathrm{Spm^{4+}}$ are governed by electrostatics and volume exclusion.

\noindent\textbf{Simulations:} 
We used OpenMM \cite{Eastman17PLOS} to simulate ssDNA  on a Graphics Processing Unit (GPU). Long-range Coulomb interactions were treated using Particle-Mesh-Ewald (PME) algorithm \cite{Darden93JCP}.  In order to enhance conformational sampling, the simulations were performed by integrating the Langevin equation in the low friction limit \cite{Honeycutt1992}. 
The equation of motion for a bead or ion, with coordinate $\vec{r_i}$, is,
\begin{equation}
m_{i} \ddot{\vec{r_i}} = -m_{i}\zeta \dot{\vec{r_i}} + \vec{F_c} + \vec{\Gamma},
\label{Eq. Langevin_dynamics}
\end{equation}
where $m_i$ is the mass of the  bead, $\zeta$ is the friction coefficient, 
$\vec{F_c} = -\frac{\partial U_{TIS-ION}(\{{\bf r}\})}{\partial {\vec{r_i}}}$, and $\vec{\Gamma}$ is the random force with a white noise spectrum.
The autocorrelation function of the random force in the discretized form is $\left < \Gamma (t) \ \Gamma (t+nh) \right > = \frac{2 \zeta m_{i}k_B T}{h} \delta_{0,n}$, where $n = 0, 1, ...$ and $\delta_{0,n}$ is the Kronecker delta function.  
The equations of motion were integrated using Langevin leap-frog method \cite{Hockney2021} using $\zeta$ = 0.01 picosecond$^{-1}$. The integration time step is 2 femtoseconds. 
We performed simulations at $T$ = 277 K.

A single ssDNA polymer is placed in a cubic box of length $L$ whose value depends on the ion concentration. To minimize finite size effects, we used periodic boundary conditions. Neutrality of the sample is maintained by adding an appropriate number of $\mathrm{Cl^{-}}$ ions. In all the simulations, 20 mM NaCl was also included.


\section*{Results}
\noindent \textbf{Calculation of $l_p$:}
We computed $l_p$  by fitting the distribution function ${P(r_{ee}, L)}$ derived for a semi-flexible chain\cite{Hyeon06JCP,Habook}, where $r_{ee}$ (=$R_{ee}/L$, $R_{ee}$ is the end-to-end distance of the chain and $L$ is the contour length) is the normalized end-to-end distance to the simulated end to end distribution.
The distribution function is given as,
\begin{equation}
\label{eq:eq_wlc_ree}
P(r_{ee},L) = \frac{4\pi A r_{ee}^{2}}{(1 - r_{ee}^2)^{\frac{9}{2}}}\exp \bigg(\frac{-3t}{4(1-r_{ee}^{2})}\bigg),
\end{equation}
where A is the normalization constant =$\bigg(\frac{4({\frac{3t}{4}})^{\frac{3}{2}}\exp(\frac{3t}{4})}{\pi^{\frac{3}{2}}(4+\frac{12}{(3t/4)} + \frac{15}{{(3t/4)}^{2}})}\bigg)$ and $t = \frac{L}{l_{p}}$. Previously \cite{Murphy2004} FRET experiments in conjunction with Eq. \ref{eq:eq_wlc_ree} were used to estimate  $l_p$ for ssDNA in monovalent cations.  

\begin{figure}[ht!]
\centering
	\includegraphics[width=0.78\textwidth]{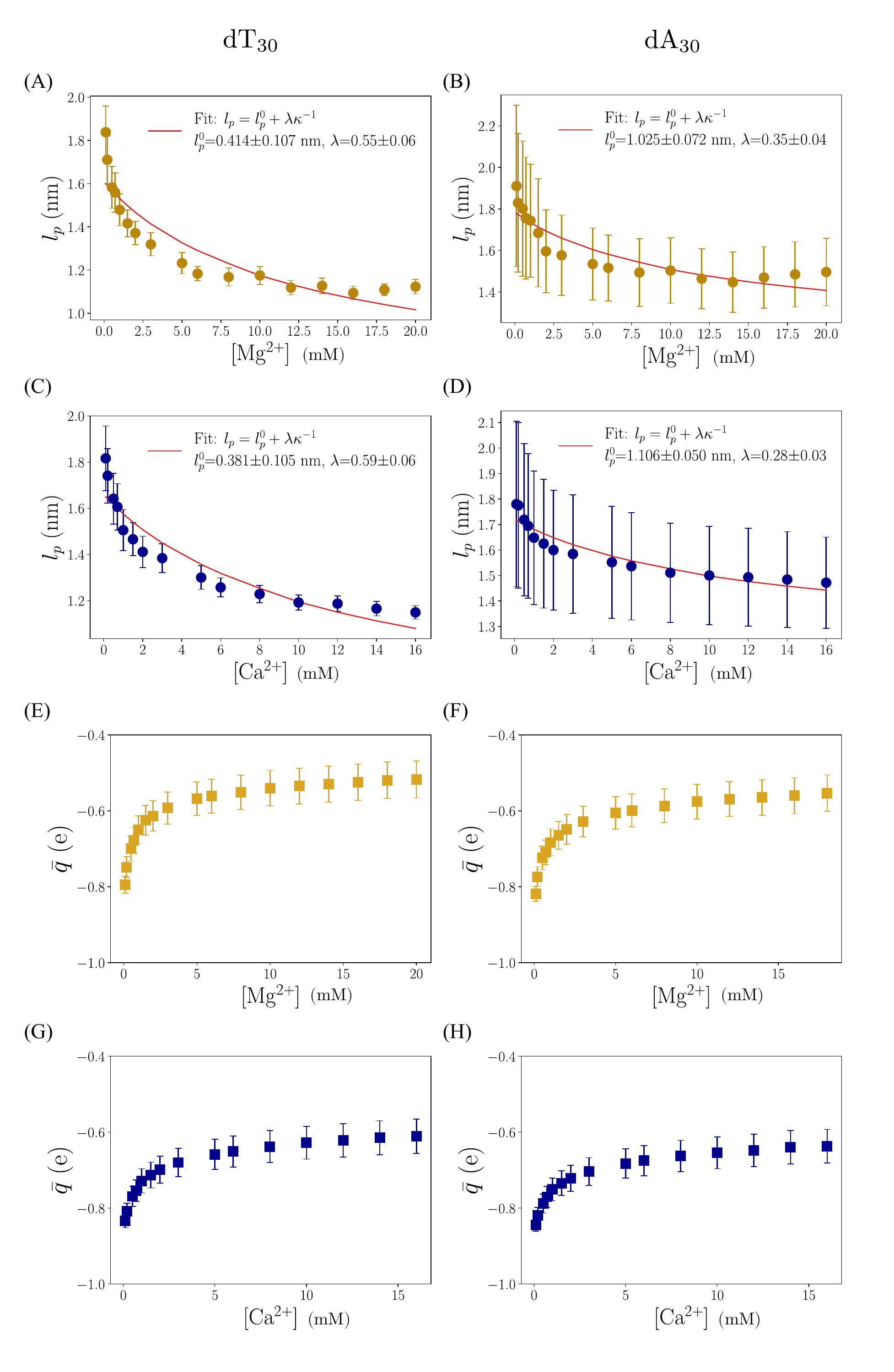}
	\caption{ {\textbf{Persistence length in divalent cations:} (A-B) Dependence of $l_p$ as a function of $\mathrm{Mg^{2+}}$ concentration, in 20 mM NaCl. (A) is for $\mathrm{dT_{30}}$ and (B) is for  $\mathrm{dA_{30}}$ (C, D) Same as (A, B), except the results are shown in $\mathrm{Ca^{2+}}$. The solid red curves are fits to the equation (Eq. \ref{leFlex}). The parameters are specified in the legends. (E)  Renormalized charge (see Eq.~\ref{eq:RenCharge}) on $P$ groups of $\mathrm{dT_{30}}$ and (F) $\mathrm{dA_{30}}$, in $\mathrm{Mg^{2+}}$.
(G, H) Same as (E, F), except the results are shown for $\mathrm{Ca^{2+}}$.\label{fig:fig_1} }}
\end{figure}

\noindent \textbf{Persistence length in  divalent cations:} Figure~\ref{fig:fig_1} shows that the persistence length
$l_p$ of ssDNA decreases monotonically as the
Mg$^{2+}$ (or Ca$^{2+}$) concentration increases.   
Because the ssDNA chains are flexible, we fit the
simulated $l_p$ to Eq. \ref{leFlex}.
The excellent fits (solid red lines in
Figure~\ref{fig:fig_1}) reveal two notable trends. First, the linear dependence of $l_p$ on
$\kappa^{-1}$ holds in the presence of divalent
cations.  These findings are noteworthy because the assumptions (PE is flexible and the electrostatic interactions are described by the DH potential)  leading to Eq. \ref{leFlex} \cite{Ha95Macro} are not usually satisfied in divalent cations.   Second, the values of $l_p^{0}$ extracted from fits reflect the sequence dependent intrinsic stiffness. The flexible $\mathrm{dT_{30}}$ ($l_p^{0} \approx$ 0.4 nm) adopts a random-coil-like conformation (Figure~\ref{fig:fig_0}D) due  to weak stacking between adjacent bases. 
In contrast, $\mathrm{dA_{30}}$, which is more helical (Figure~\ref{fig:fig_0}D) as a result of 
stronger base stacking, has a larger $l_p^{0}$ ($\approx$ 1.1 nm). In both Mg$^{2+}$ and Ca$^{2+}$ the values of $l_p^{0}$ are sequence independent. 

A note on relation to experiments is worth making.  The $l_{p}^{0}$ values obtained by fitting the simulated $l_p$ fall well within the range of experimentally reported $l_{p}^{0}$ values (0.55 nm to 1.5 nm) for different ssDNA constructs, mostly in monovalent salts \cite{Tinland1997,Kuznetsov2001,Woodside2006,Saleh2009,McIntosh2011,Chen12PNAS,Alemany2014,Bosco2014,Joan2016,Jacobson2017,Murphy2004}. Interestingly, $l_p$ extracted from force extension traces generated in single molecule pulling experiments \cite{Bosco2014} on long (6770 base pairs) ssDNA sequence for a limited number of $\mathrm{Mg^{2+}}$ data points was fit using Eq. \ref{leFlex} with $l_p^{0} \approx$ 0.7 nm. In light of the present work, it would be interesting to do additional experiments on ssDNA homopolymers in $\mathrm{Mg^{2+}}$ and $\mathrm{Ca^{2+}}$.


\noindent \textbf{Charge renormalization:} The reduction in $l_p \sim c^{-0.5}$   occurs in two steps. There is a substantial  reduction in $l_p$ upon addition of  small amount ($\sim$ 0.1 mM) of $\mathrm{Mg^{2+}}$  and $\mathrm{Ca^{2+}}$ ions  (of the order of 0.1 mM) that is followed by a moderate change in $l_p$ at higher concentrations. The decrease in $l_p$ could be qualitatively explained using counter ion condensation theory \cite{OosawaBook,Manning69JCP}. The simulations could be used to directly calculate the number of monovalent and divalent ions condensed along the phosphate backbone, which would enable the calculation of the renormalized charge $\bar{q}$ per phosphate bead. We assume that  a counter ion is condensed onto  a phosphate if the ion is within a cutoff distance $r_c = (r_{P} + r_{i} + \Delta r )$, where $r_{i}$ is the radius of $i^{th}$ bead and $\Delta r$  taken to be 0.15 nm.
The renormalized charge is computed using, 
\begin{equation}
\bar{q} =\frac{1}{N}\sum_{i=1}^{N} q_{P}^{i} + Z_{1}f^{i}_{1} + Z_{2}f^{i}_{2},
\label{eq:RenCharge}
\end{equation} 
where $q_{P}^{i}$ is the bare charge on the $i^{th}$ phosphate group (-1), $f^{i}_{1}$ is the number of monovalent and $f^{i}_{2}$ is the number of divalent ions bound to $i^{th}$ phosphate, and $Z_{1}$ and $Z_{2}$ are the valency of the monovalent and divalent ions respectively.  

In 20 mM NaCl solution, with no added $\mathrm{Mg^{2+}}$ ions, $\bar{q}e$ is close to -0.9 for both the ssDNA constructs. Addition of $\mathrm{Mg^{2+}}$ ions up to 1 mM accounts for $\approx$ 25 $\%$ charge reduction while further addition of ions up to 20 mM led to only $\approx$ 13 $\%$ charge reduction. The value of $\bar{q}e$ saturates  at $\approx$ 50 $\%$ charge reduction. In agreement with counter ion condensation theory, majority of the charge neutralization is due to condensed $\mathrm{Mg^{2+}}$ ion ($\approx$ 47 $\%$ for $\mathrm{dT_{30}}$ at highest concentrations) while $\mathrm{Na^{+}}$ neutralizes $\approx$ 1 $\%$ of phosphate charges.

Following Oosawa-Manning theory\cite{OosawaBook,Manning1969}, we  computed the renormalized charge on the phosphate using, $q_{OM} = \frac{b}{l_{B}(T)}$ where $l_{B}(T)$ is the Bjerrum length and $b$ is the length per unit charge taken as 0.44 nm for single-stranded polynucleotides \cite{Olson1976}. We estimate that $q_{OM}$  is -0.6$e$, while from our simulations we obtain a value close to -0.5$e$ for $\mathrm{dT_{30}}$ and -0.6$e$ for $\mathrm{dA_{30}}$. It follows that  $\mathrm{Mg^{2+}}$  is more efficient in neutralizing charge on $\mathrm{dT_{30}}$  compared to $\mathrm{dA_{30}}$.

\begin{figure}[ht!]
\centering
	\includegraphics[width=0.9\textwidth]{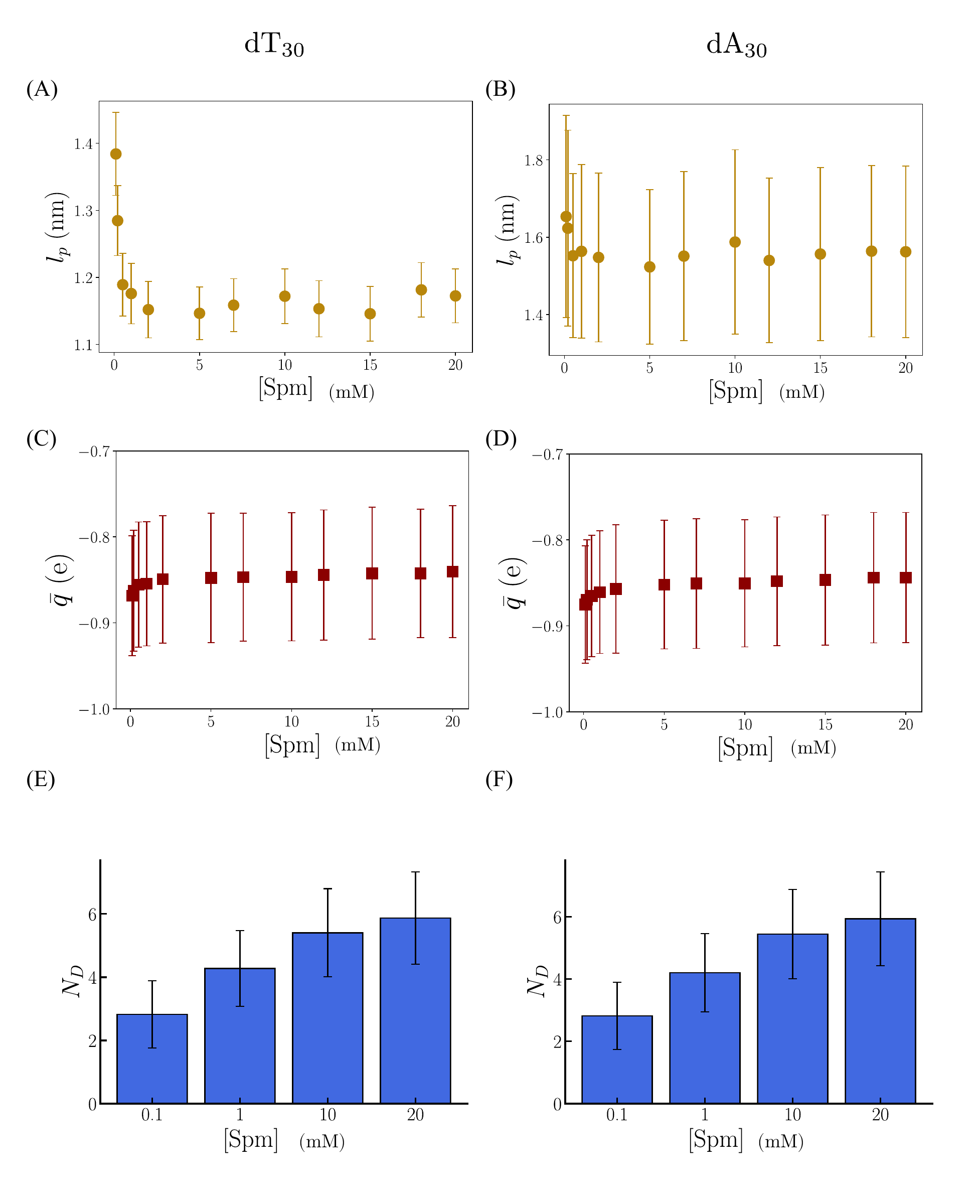}
	\caption{\label{fig:fig_2} 
\textbf{Persistence length in spermine:} (A) Dependence of $l_p$ of $\mathrm{dT_{30}}$ and (B) $\mathrm{dA_{30}}$ as a function of $\mathrm{Spm}$ concentration in 20mM NaCl. (C) Renormalized charge (see Eq.~\ref{eq:RenCharge}) on $P$ groups of $\mathrm{dT_{30}}$ and (D) $\mathrm{dA_{30}}$.
(E) Number of $\mathrm{Spm}$ molecules ($N_{D}$) in direct contact with at least one of the phosphate ($P$) groups, for $\mathrm{dT_{30}}$ and (F) $\mathrm{dA_{30}}$. A direct contact means at least one of the amine moiety in $\mathrm{Spm}$ and the $P$ groups  are  within  the Bjerrum length.
}
\end{figure}

\noindent\textbf{Persistence length in spermine:}
We performed simulations to estimate the dependence of $l_p$ on spermine concentration, which was varied from 0.1 mM to 20 mM.
Interestingly, $l_p$ does not change beyond $0.5\,\mathrm{mM}$ (Figure~\ref{fig:fig_2}) in both $\mathrm{dT_{30}}$ and $\mathrm{dA_{30}}$. The lack of change in $l_p$ even when the $\mathrm{Spm^{4+}}$ is increased by nearly two orders of magnitude (0.1 mM to 20 mM) is striking.
At all concentrations, $l_p$ is larger for $\mathrm{dA_{30}}$ than for $\mathrm{dT_{30}}$. In the absence of theory to account for the anisotropic nature of spermine molecule with four charges, there is no clear explanation of the simulation results.


The lack of change in $l_p$ is also reflected in the renormalized charge $\bar{q}$ (Eq.~\ref{eq:RenCharge}). It also saturates near $[\mathrm{Spm}] \approx 0.5\,\mathrm{mM}$ (Figures~\ref{fig:fig_2}C,D).
At all concentrations, $\bar{q}$ remains lower than for $\mathrm{Mg^{2+}}$ and $\mathrm{Ca^{2+}}$, indicating weaker charge screening by Spm.

These trends point to important differences between spermine and divalent cations.  At neutral pH, spermine has four protonated amine groups linked along a flexible backbone, unlike spherical divalent cations such as $\mathrm{Mg^{2+}}$ and $\mathrm{Ca^{2+}}$.
This connectivity produces anisotropic interactions with the nucleotides and enhances the effects of excluded volume. 
A bound $\mathrm{Spm^{4+}}$  could  sterically hinder  neighboring phosphate sites, limiting the number of spermine molecules that bind to ssDNA simultaneously. To illustrate the importance of spermine volume exclusion, we define a direct Spm--DNA contact when at least one amine group is within $l_{B}$ ($\approx 0.7\,\mathrm{nm}$) of a phosphate group. Let $N_D$ be the total number of such bound molecules.
Figures~\ref{fig:fig_2}E and F show that $N_D$ remains nearly constant (3-6) over a wide range of spermine concentrations (0.1 mM-20 mM). This is consistent with steric saturation of binding sites, described above. Once this saturation is reached, further increases in spermine concentration do not change the ssDNA conformation, hence the $l_p$.

\begin{figure}[ht!]
\centering
	\includegraphics[width=
    \textwidth]{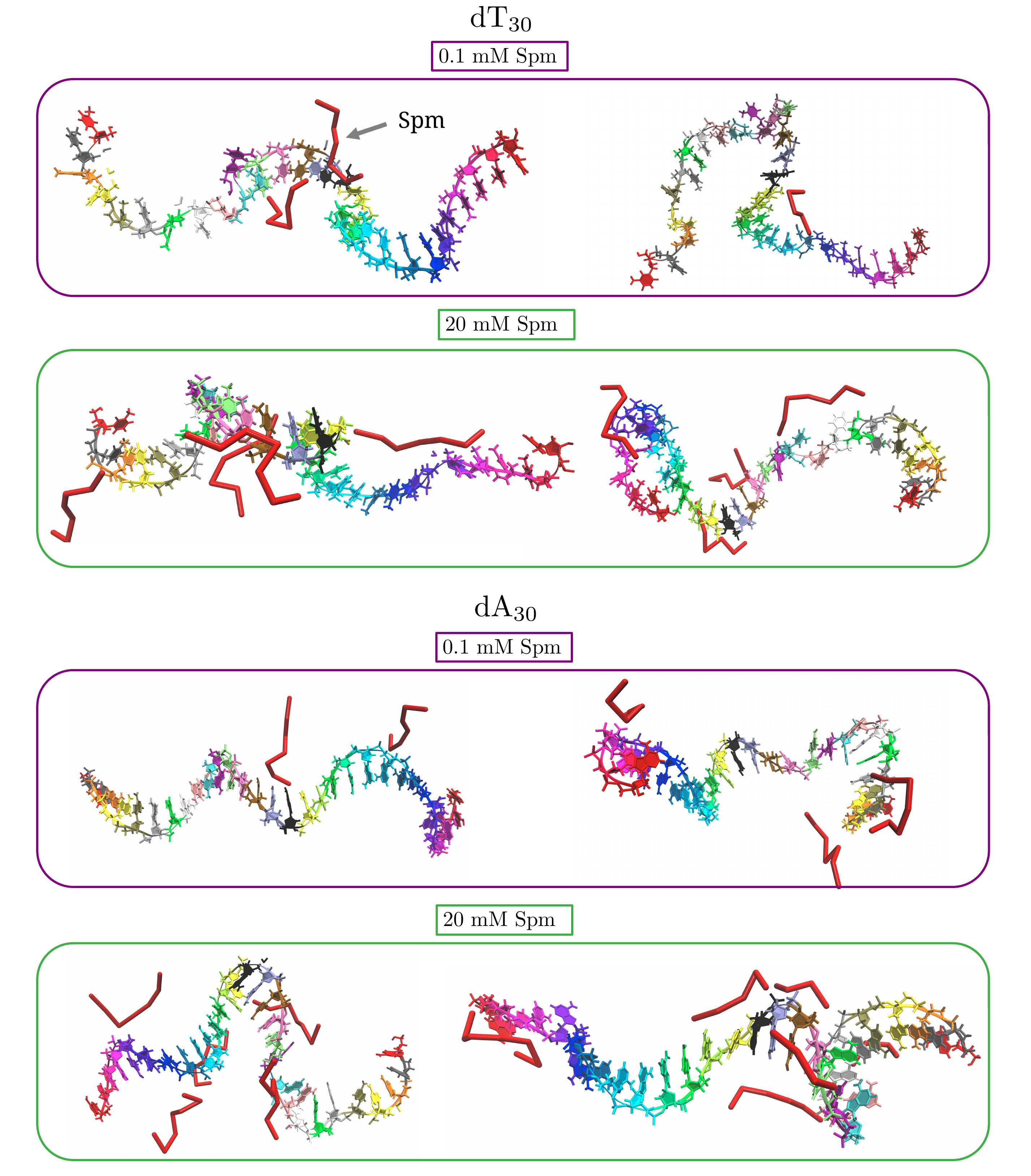}
	\caption{\label{fig:fig_3} 
\textbf{ssDNA conformations in spermine:} Reconstructed all-atom structures for $\mathrm{dT_{30}}$ and $\mathrm{dA_{30}}$, obtained from coarse-grained simulations in 0.1 mM and 20 mM Spm. To aid visualization, nucleotides
are assigned different colors based on the sequence number. Bound spermine molecules are shown in red.
}
\end{figure}

{Figure~\ref{fig:fig_3}} shows representative all-atom conformations of $\mathrm{dT_{30}}$ and $\mathrm{dA_{30}}$ with bound Spm molecules in 0.1 mM and 20 mM Spm.  It is clear that $N_D$ changes only marginally despite a 200-fold increase in Spm concentration, which is consistent with the results in Figures~\ref{fig:fig_2}E and F.
The all-atom structures were generated by backmapping the coarse-grained structures. The details of this procedure are described elsewhere \cite{Mondal24JCTC}.

To our knowledge, there are no experiments on the persistence length as a function of spermine concentration. Our results thus offer a specific prediction for future tests. Single-molecule force spectroscopy or FRET-based end-to-end distance measurements, similar to those used for divalent cations, could test our predictions.  Such experiments would also elucidate  whether steric saturation of binding sites is a general feature of polyamine-ssDNA interactions.

\medskip
\medskip
\noindent \section*{Conclusions}  The persistence length of short ssDNA chains ($\mathrm{dT_{30}}$ and $\mathrm{dA_{30}}$) in polyvalent cations was calculated using simulations of the coarse-grained TIS-DNA model.
In divalent cations, the electrostatic persistence length ($l_e$) is proportional to the Debye length, which  unexpectedly agrees with theories developed for flexible polyelectrolytes. The bare persistence length is sequence dependent. The variation of $l_p \sim \kappa^{-1} \sim c^{-0.5}$ is explained by charge renormalization on the phosphate groups due to counterion condensation. 

In contrast, we find that $l_p$  is almost independent of $\mathrm{Spm^{4+}}$ concentration. There is no theoretical explanation for this unexpected prediction. 
It should be pointed out that binding of $\mathrm{Spm^{4+}}$ to the phosphate groups results is favorable. 
However, it is accompanied by a loss in translational and rotational entropy of spermine. Importantly,  the number of $\mathrm{Spm^{4+}}$ that can bind to ssDNA is restricted because of the large size of the cation. These factors have to be taken into account to develop  a reasonable theory. Given the importance of polyamines in general  and spermine in particular, experiments are needed to validate our predictions.

\bigskip
\noindent \textbf{Acknowledgments:} This article is a tribute to M. Muthukumar whose numerous works in polymer physics have been a source of inspiration to DT. We are grateful to NSF (grant CHE 2320256) and  the Welch Foundation (grant F-0019), administered through the Collie-Welch Regents Chair, for supporting this research. We thank the Texas Advanced Computing Center and Pittsburgh Supercomputing Center for providing generous computational resources.

\bigskip

\noindent \textbf{Data Availability:} The data that supports the findings of this study are available from the corresponding author upon reasonable request.

\newpage

 \clearpage
 \bigskip

\clearpage

\newpage
\bibliography{ssNA}
\bibliographystyle{unsrt}

\end{document}